# A Conversation with Dorothy Gilford

Edward J. Wegman and Wendy L. Martinez

*Abstract.* In 1946, Public Law 588 of the 79th Congress established the Office of Naval Research (ONR). Its mission was to plan, foster and encourage scientific research in support of Naval problems. The establishment of ONR predates the National Science Foundation and initiated the refocusing of scientific infrastructure in the United States following World War II. At the time, ONR was the only source for federal support of basic research in the United States. Dorothy Gilford was one of the first Heads of the Probability and Statistics program at the Office of Naval Research (1955 to 1962), and she went on to serve as Director of the Mathematical Sciences Division (1962 to 1968). During her time at ONR, Dorothy influenced many areas of statistics and mathematics and was ahead of her time in promoting interdisciplinary projects. Dorothy continued her career at the National Center for Education Statistics (1969 to 1974). She was active in starting international comparisons of education outcomes in different countries, which has influenced educational policy in the United States. Dorothy went on to serve in many capacities at the National Academy of Sciences, including Director of Human Resources Studies (1975 to 1978), Senior Statistician on the Committee on National Statistics (1978 to 1988) and Director of the Board on International Comparative Studies in Education (1988 to 1994). The following is a conversation we had with Dorothy Gilford in March of 2004. We found her to be an interesting person and a remarkable statistician. We hope you agree.

*Key words and phrases:* Office of Naval Research, National Academy of Sciences, National Center for Education Statistics, Committee on National Statistics, Board on International Comparative Studies in Education.

*Edward J. Wegman is Professor, Center for Computational Statistics, George Mason University, MS 6A2, Fairfax, Virginia 22030, USA e-mail: ewegman@gmu.edu. Wendy L. Martinez is Program Officer, Office of Naval Research, One Liberty Center, 875 North Randolph Street, Room 1177, Arlington, Virginia 22203-1995, USA e-mail: martinwe@onr.navy.mil.*

## EARLY LIFE

**Ed:** This is Edward Wegman and I am here with Wendy Martinez in the home of Dorothy Gilford. Dorothy, why don't you tell us about your early history, where you grew up and what sort of background your family had?

**Dorothy:** I was born in 1919. I just celebrated my 85th birthday. I was born in Ottumwa, Iowa, but I didn't live there very long. My father worked for Kelly-Springfield Tire Corporation. He was moved to Lincoln, Nebraska when I was 2. We moved to Los Angeles when I was 5. When I was 8, my father was sent to Seattle, where he was the branch manager for the region. I really grew up in Seattle, where I went to grade school and high school. Then

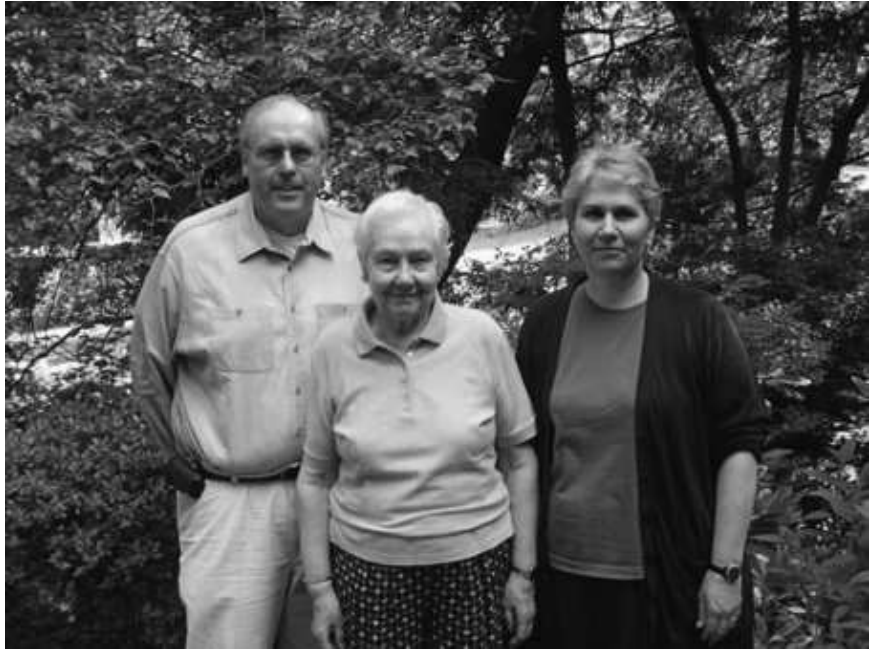

Fig. 1. *This photo was taken in May 2006. From left to right: Edward Wegman, Dorothy Gilford and Wendy Martinez. Dorothy was the second Director of the Statistics and Probability Branch at the Office of Naval Research (ONR), Ed was the sixth Director, and Wendy is the current one.*

I went to the University of Washington to major in mathematics. Incidentally, I majored in mathematics because I asked the guidance counselor for some advice when I left high school. She said it was very rare for a woman to have good grades in mathematics, and I should major in mathematics. That really wasn't the best advice, considering the career opportunities at that time.

**Wendy:** She must have felt that you would have gotten good grades at the university.

**Dorothy:** Yes, but if you looked a little beyond the university, the career opportunities were very, very limited at that time.

**Wendy:** What year was that?

**Dorothy:** 1935.

**Ed:** Was that high school or college?

**Dorothy:** High school. I did a year of "post-graduate high school" because I was only 15 when I graduated. I took mathematics at the University of Washington, and I minored in botany. I liked botany very much. I took all of the mathematics courses. In my senior year, the only woman faculty member in the department, Mary Haller, was the professor that *all* the engineering students liked. As a matter of fact, the engineering department sent them to her, because she was the best teacher. She never made full professor because she didn't publish. She advised me that mathematics was not a good career for women, and she urged me to do something else, but I decided I would go ahead and do some graduate work in mathematics. I was lucky in my first year of graduate work. Z. W. Birnbaum[1] was on the faculty. I took his course in what was actuarial mathematics, but then he got into statistics. I took a course in statistics and probability from him, and I really liked it. I liked it very, very much. So, I decided I would shift a little bit out of pure mathematics into statistics.

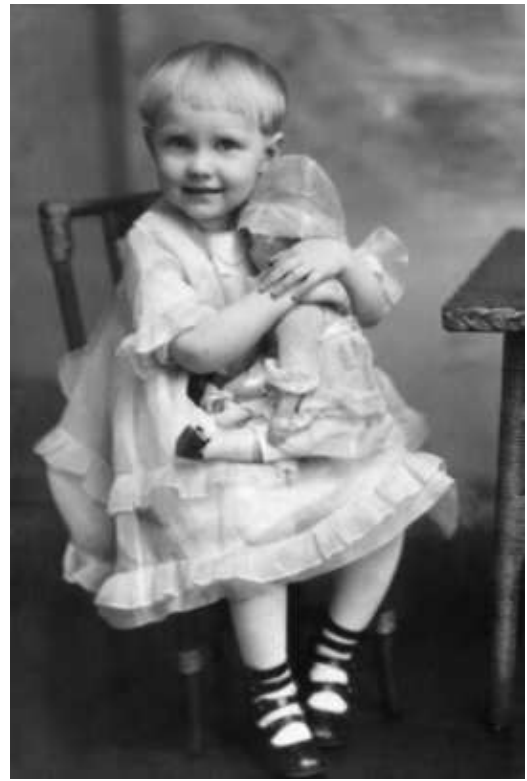

Fig. 2. *This shows Dorothy as a young child of two or three years old.*

**Ed:** So, as an undergraduate, what sort of mathematics were you focused on?

**Dorothy:** Theoretical mathematics.

**Ed:** Such as abstract algebra and topology?

**Dorothy:** No, not topology—Boolean algebra, real analysis, differential equations.

**Wendy:** Did you find that your parents were supportive of your college endeavors?

**Dorothy:** My father died when I was only 8. My mother was very supportive of my education. She did everything she could to see that I had a good education, including moving. When I got to junior high, we moved to an apartment building across the street from John Marshall Junior High School. When I went to the university, we moved to the university district. She did everything she could to make it easy for me. Actually, we did not have very much money. I worked in the university library half time all through my university years. That was a good education, too.

## GRADUATE SCHOOL YEARS

**Wendy:** So often, I think that women had trouble going to college or doing graduate work because the families didn't support them. It sounds like you did not have those problems.

**Dorothy:** No, not at that point. When I got my Master's degree, I decided I wanted to go on for a Ph.D. I applied for fellowships at eight schools. It was during the war, and there weren't many men around. So, I was awarded a fellowship at each of the eight institutions. I had no competition! I chose to go to Bryn Mawr. I always wanted to go there, so

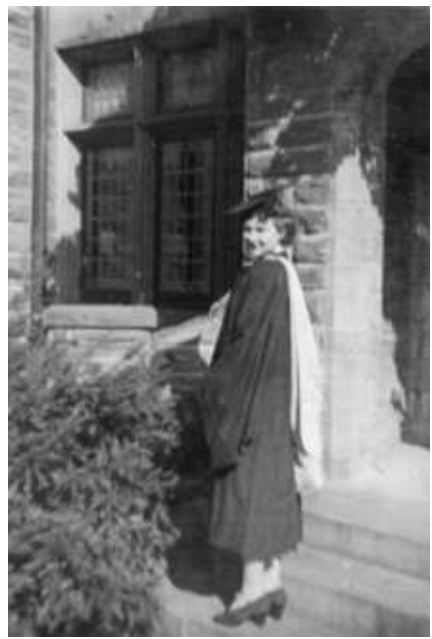

FIG. 3. *This shows Dorothy Gilford at Bryn Mawr graduation ceremonies in 1943.*

I went there for graduate school. I did some work with Hilda Geiringer[2] on some genetics problems that could have gone on to be a dissertation. But, I went to New York for a meeting of the IMS (Institute of Mathematical Statistics) where I met Harold Hotelling.[3] He offered me a fellowship at Columbia. I thought about it, and I decided I would go to Columbia. I spent one year there. I found it a very *cold* place.

**Wendy:** Wasn't there a big statistics group at Columbia?

**Dorothy:** Yes, it was during and after World War II. I was at Columbia during the war, 1942–1943. Abraham Wald worked there. There was a Statistical Research Group, and Harold Hotelling, Jacob Wolfowitz, and Al Bowker were all in that group.[4] At Columbia, I had courses with Wald and Hotelling

---

[1] Z. W. Birnbaum (1903–2000) was born in Austria-Hungary and studied mathematics and actuarial science. He came to New York in the late 1930s, where Harold Hotelling told him about a job at the University of Washington. This led to a distinguished career of over 60 years in the Seattle area.

[2] Hilda Geiringer von Mises (1893–1973) was born in Hungary and received her doctorate in mathematics. She came to the United States in 1939 and was appointed as a lecturer at Bryn Mawr College. She had a distinguished career as a university professor of probability and statistics.

[3] Harold Hotelling (1895–1973) is well known for his many contributions to statistics. He was a faculty member at Columbia University from 1931 to 1946.

[4] Abraham Wald (1902–1950) is known for his contributions to statistical decision theory and sequential analysis, much of which was done to support the military in World War II. He joined the faculty at Columbia University in 1941 and

in mathematics, some really famous people. I didn't like Columbia. The faculty kept office hours two hours a week, and when you went they were not in the office, so you never could find out anything.

**Ed:** So, you went back to Bryn Mawr?

**Dorothy:** Yes, I went back to Bryn Mawr for a teaching assistant position. I also taught a course in statistics there and did some more course work. At the end of that year, there was an announcement on the bulletin board for a position at George Washington University for an assistant professor. I came down to Washington, and I got the job!

**Ed:** So, that was pre–Ph.D.?

**Dorothy:** Yes, that was pre–Ph.D. I taught there for two years, and it was a grueling schedule. My first semester, I taught three graduate courses and one undergraduate course. The associate professor, who was my colleague, was teaching one graduate course and one undergraduate course, and that didn't seem quite right to me. I decided that I didn't want to stay. I talked to Hotelling; I was doing my dissertation with Hotelling. He said he would see if I could come down to North Carolina. He had moved to North Carolina, Chapel Hill, by that time.[5] I was there for a year working on my dissertation. I finished it and went up to Columbia. (I hadn't taken any classes formally at Chapel Hill, so I couldn't get a degree there.) I went up to Columbia with my dissertation in hand. Wald looked at me, and he said, "Well, you've written a multivariate analysis distribution of $t_0$-squared and that is not my field. Anderson, whose field it is, is in Stockholm. Why don't you come back in a year with your dissertation?" I was devastated.

**Ed:** So, this is Ted Anderson?

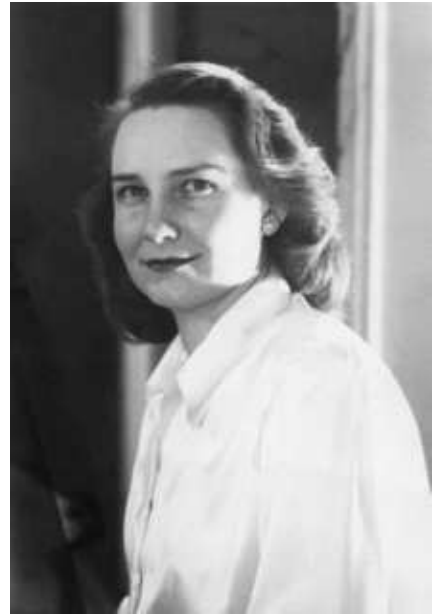

Fig. 4. *This is Dorothy Gilford in 1947.*

**Dorothy:** Yes, Ted Anderson. Wald was the acting head of the department at that time. Hotelling had been the head, but Hotelling went to North Carolina. That is why I was down there for a year working on it.

**Ed:** Hotelling went to North Carolina around 1946.

**Dorothy:** Yes, Gertrude Cox talked Hotelling into coming to Chapel Hill.

**Wendy:** Did you know Gertrude Cox?

**Dorothy:** Oh yes, we were good friends.

**Wendy:** So, when did you get your Ph.D., then?

**Dorothy:** I didn't get it. I got married!

**Wendy:** When did you get married?

**Dorothy:** I got married in 1950. I have a dissertation, which is referenced in Ted Anderson's multivariate analysis book because I was the first person to get that distribution, but I didn't get the Ph.D. I went back to Washington. I did some consulting while I was there and consulted at the Navy Medical Research Institute. One of the people there moved to what was then the Civil Aeronautics Administration, and he called me and asked me if I would like to come and work there. I went for an interview, and I soon became the Chief of the Biometrics Branch in the Medical Division of the Civil Aeronautics Administration. This was a very interesting appointment. It involved doing analysis of the reports of physical examinations of private and commercial pilots. While I was there, someone else I knew told me about a job at the Federal Trade Commission and said I should go interview for that, since

---

remained there until he died in a plane crash while visiting India.

Jacob Wolfowitz (1910–1981) was born in Poland and emigrated to the United States in 1920. He began his career as a high-school mathematics teacher. After his Ph.D., he went on to become a professor of mathematics at Cornell University.

Albert Bowker (1919–) received his Ph.D. in statistics from Stanford University, under Jacob Wolfowitz. He was at the Columbia Statistical Research Group from 1943 to 1945. He was the founding Chair of Stanford's Statistics Department and went on to serve as Chancellor of the University of California, Berkeley.

[5]The Department of Statistics at the University of North Carolina at Chapel Hill was organized by the vision of Gertrude Cox in 1946 with Harold Hotelling as its first Chairman.

it would be a promotion. I did, and I got the job. I was Deputy Director of the Division of Financial Statistics, which produced a quarterly report on finances of manufacturing corporations. It was a joint product of the Federal Trade Commission and the Securities and Exchange Commission. They did the registered corporations, and I did a survey of all the others. I did that for two or three years. I got another phone call, this time from Herb Solomon,[6] who said, "Why don't you come over to ONR? There is going to be a job here." I did that too. Each time I was getting a promotion. Joe Weyl was the Director of the Mathematical Sciences Division at the time.

## YEARS AT THE OFFICE OF NAVAL RESEARCH

**Ed:** When did you go to ONR?
**Dorothy:** I went there in 1955.
**Wendy:** Can you tell us more about Joe Weyl?
**Dorothy:** He was the son of the famous Hermann Weyl, a mathematician at Princeton.[7] I ran the Statistics Branch for several years.
**Ed:** So, Herb had left ONR by that time?
**Dorothy:** Yes, he was leaving to go to California. I was replacing Herb Solomon as the second Head of the Statistics Branch. I enjoyed ONR. I enjoyed meeting all of the top people in the field and going to the professional society meetings. During that time I was secretary of IMS, so I went to a lot of meetings. Then Joe Weyl moved up to be the Technical Director.[8] Rather than bringing in another person to head the division, they moved Fred Rigby, who headed the Logistics Branch, to that position. Then they combined the Statistics and Logistics Branches, so I was heading both Statistics and Logistics.
**Ed:** So, was the *Naval Research Logistics Quarterly*[9] in business then?
**Dorothy:** Yes, it was.
**Ed:** So, the Navy's interest in logistics, as I understand it, was partly in response to the Nautilus

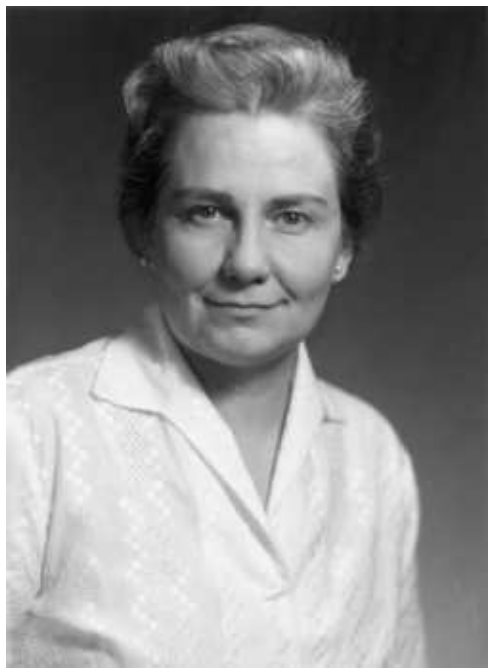

FIG. 5. *Here is Dorothy Gilford in 1960 during her time at ONR.*

nuclear submarine. There was a lot of interest, as far as I know, in having to deal with submarines that were going to be deployed for six months or more.
**Dorothy:** Yes. At ONR Fred Rigby was promoted to Division Director, and later he went to Texas Tech as a Dean. When he left, I was promoted to be Division Director.
**Wendy:** Were you the first female Division Director at ONR?
**Dorothy:** No, Mina Rees was the first female Division Director. She was there at the founding of ONR. Joe Weyl, Fred Rigby and I followed her.
**Wendy:** Could you tell us a little more about Mina Rees and Herb Solomon?
**Dorothy:** Mina Rees was an interesting woman. When she left ONR, she became Dean of Hunter College, which is the women's college associated with the City University of New York. That was a very nice job for her. She was a mathematician and very prominent in mathematics. She was a very nice, genuine person. I met with her after I went to NCES (National Center for Education Statistics). She wanted to talk about ONR, and we had lunch together. Herb Solomon I knew very well. We were graduate students at Columbia at the same time.
**Ed:** You were both graduate students there?
**Dorothy:** Yes, as was Al Bowker. Herb Solomon was eventually funded by ONR when he was at Stan-

---

[6]Herb Solomon was the founding Director of the Statistics Branch at the Office of Naval Research, where he served between 1948 and 1955. Herb died September 20, 2004.

[7]Hermann Weyl was Visiting Professor at Princeton in 1928–1929. He fled Germany in 1933 to work at the Institute for Advanced Study in Princeton, where he remained until his retirement in 1952. He died on December 8, 1955.

[8]The Technical Director was the highest-ranking civilian position in science and technology at ONR.

[9]Volume 1, Number 1 of the *Naval Research Logistics Quarterly* was published in March 1954.

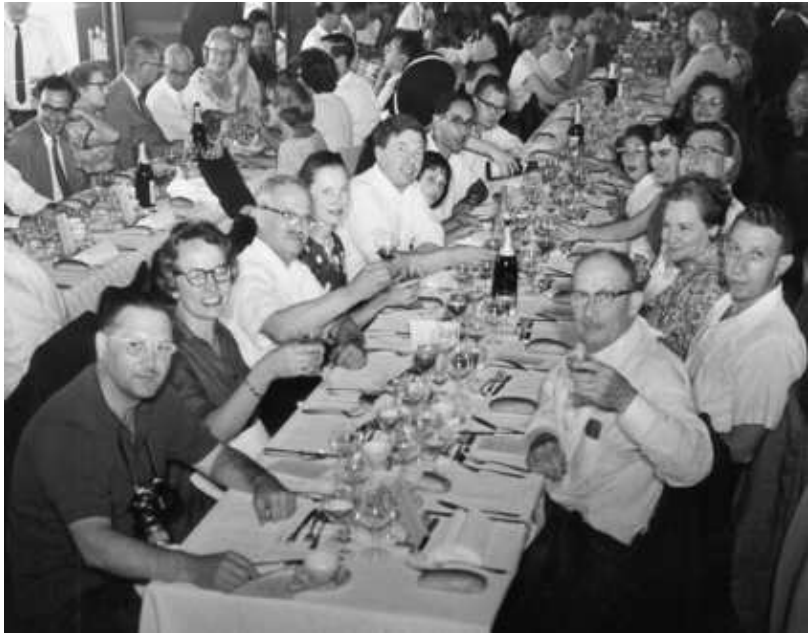

FIG. 6. *Here is a picture of Dorothy with colleagues at the 1961 ISI meeting in Paris. On the right side, W. J. Youden, Ingram Olkin, Dorothy Gilford, Emanuel Parzen, Carol Parzen, Ellen Chernoff, Judy Chernoff. On the left side, starting with the second person, Betty Scott, Jerzy Neyman, Ann Durbin, Jim Durbin, Miriam Chernoff, Herman Chernoff (www.stat.tamu.edu/events/parzenprize/pictures/index.html).*

ford. Al, Herb and Jerry Lieberman[10] were founders of the Statistics Department at Stanford.

**Ed:** So, Mina Rees was the founding Division Director for Mathematics?

**Dorothy:** She was not only the Division Director; she was one of the three or four key founders of ONR, recognizing the need for it and getting it started. Then she was the Division Director.

**Wendy:** So she probably made the case to Congress then?

**Dorothy:** I'm sure she was involved in that.

**Wendy:** In what division was the Statistics and Logistics Branch?

**Dorothy:** It was in the Mathematical Sciences Division.[11] Altogether, I was at ONR for 12 years. I was Division Director for six years. I left in 1968.

That was a challenging position, because I really had to learn a lot to go to program council[12] and justify projects in aerodynamics, structural mechanics and information sciences. I found information sciences especially interesting.

**Wendy:** So, they had program council back then?

**Dorothy:** Yes, I can tell you a funny story about that. Because I was a Division Director, I went to the Admiral's staff meeting. I was the only woman Division Director. At the first meeting I attended, I went in and sat down by the man who was the special assistant to the Admiral. He was a person I knew fairly well. The Admiral came in and everyone stood up, so I did, too. We sat back down. The Admiral looked around for about a minute. Then he started the meeting. Afterwards, I asked my friend why they had that silence. He said it was customary to tell an off-color joke at the beginning, and the Admiral didn't think that was proper with a woman present. I suggested I could go five minutes late or something like that, but that was the end of the jokes. I liked that job very much. I was also very active in the International Statistical Institute (ISI) at that time, too.

---

[10]Gerald J. Lieberman (1925–1999) was born in New York. He obtained a master's degree in statistics from Columbia University and went to Stanford as a doctoral student in statistics. He co-authored two influential books: *Handbook of Industrial Statistics* (with Al Bowker) and *Introduction to Operations Research* (with Frederick S. Hillier).

[11]ONR is organized into Programs (Branches), Divisions and Departments. Computer science was set up as a Branch within the Division sometime prior to 1978. In 1978, the Division was called the Mathematical and Information Sciences Division.

[12]Program council is a weekly meeting of Division Directors, Department Heads, Contract Officers and finance representatives. They serve in an oversight role, and they look at all new start efforts to ensure that the work is relevant and the performers are of high quality.

**Wendy:** Do you have any other funny stories about your time at ONR?

**Dorothy:** Yes, I do. I once had a contractor who was a famous statistician. He was interviewed for an article, and he said that he did his best research on the beaches in Rio. It was also published in the paper that he did his research under contract to the Office of Naval Research. I was called to the Admiral's office about funding this man who claimed he did his research on the beaches. Our contractor was a very prominent statistician and a very able researcher. I tried to explain to the Admiral that many times you did get your best ideas in the shower or places outside of an office. Something is on the back of your mind all the time and then it comes together. The researcher really might be more successful when he was relaxed on the beach. The Admiral said all right, and he let it go. So, two months later the researcher's contract came up for renewal, and I renewed it! I was called to the Admiral's office again. I said, "Well, he does beautiful research. He has made some major contributions to the field." The Admiral just shook his head and let me go.

**Ed:** So, you were an ISI member?

**Dorothy:** Oh yes. I was asked to serve on the board, but I didn't really have the time to do it. In retrospect, I wish I had. During this period of time, I decided I should work on my dissertation. The Office of Naval Research sent me to Carnegie Mellon for a year.

**Wendy:** What was the subject of your second dissertation?

**Dorothy:** It was a problem in program budgeting. I was studying the relation between the stated goals of the university (which were for basic research, teaching quality and community service) and the goals of the faculty. I interviewed a large number of faculty members, asking about their goals and the actual time they were spending on these stated areas. I was matching that to the institutional goals. It was less mathematical. I did another dissertation, but I never got the degree.

**Wendy:** Did you find that the lack of an official Ph.D. limited you in your career options?

**Dorothy:** No, it did not limit me. If I had wanted to go to a university after retirement, I might have had a different problem. However, I was still offered some university positions.

**Ed:** Where did you go after ONR?

**Dorothy:** I was offered a job at the Department of Health, Education and Welfare. Alex Mood was there as the Director of the National Center for Education Statistics and, as the Assistant Commissioner for Education Statistics, he was also the advisor to the Commissioner, Office of Education. I liked ONR so much that I turned down the offer. Alex called me again in a month and said to come back and talk again. I did and talked to the Commissioner and Deputy Commissioner. Because ONR is a really special place to work, I turned it down again. Finally, they called a third time. In the interim, Congress was reviewing the ONR budget. I don't know whether it was a level budget or a cut in funding, but a congressman said, "The honeymoon with basic research is over. We are going to have a long, steady marriage." They meant it. That's when the drive started for military-oriented research.

**Ed:** That was 1968, with the Mansfield Amendment.[13]

**Dorothy:** I think it probably was. I didn't like that very much. Nonetheless, I took it to heart—that every contract had to be military oriented. I decided that I would not recommend funding a project unless it had a statement of what the military relevance was to the Navy. My Branch Heads were furious with me, especially those in more abstract fields like pure mathematics, but I insisted. So, every single one of the PR's[14] in the Mathematics Division had military relevance in it. The next year, the OMB (Office of Management and Budget) really clamped down on the budget. The Physics Division (I was gone by then) had their budget cut, but the Mathematics Division did not because everything in the division was relevant to the military according to the procurement papers. I was very proud of that. Anyway, when I was offered the job at the Department of Health, Education and Welfare for the third time, I took it.

**Wendy:** That is very interesting. I think you showed a lot of foresight.

---

[13]In the late 1960s, Senator Mike Mansfield (D-Montana) proposed an amendment that prohibited the Department of Defense from spending money "to carry out any research project or study unless such project or study has a direct and apparent relationship to a specific military function." See www.nsf.gov/nsb/documents/2000/nsb00215/nsb50/1970/mansfield.htm

[14]A PR is a procurement request. This is the document that the Scientific Officers write to initiate a contract or grant negotiation.

**Dorothy:** I was just responding to Congress. If they were going to say we had to have military relevance, even if we had to stretch it some, we could do it. For example, look at matrix algebra. When that was invented, who would have thought that would have military relevance? But, of course, it does. There are a lot of examples like that.

**Wendy:** Could you give us some examples of projects that you supported when you were Branch Head or Division Director that later became well known? For example, Ed funded some of the initial work on CART (Classification and Regression Trees) when he was the Probability and Statistics Branch Head, and now twenty years later that methodology is one of the fundamental tools in a data mining toolbox. Are there examples of that? Part of the problem with basic research is that you do not hear about the benefits of it for many years, and by that time the connection is lost.

**Dorothy:** Well, one example is the work done at the University of Washington on the capture-recapture technique, which is widely used to estimate the population size of mobile animals. Some of the books that were published back then by ONR principal investigators have had a major impact. For example, Wald's book on sequential analysis is a seminal work, and the Bowker and Lieberman book on engineering statistics was widely used.

**Ed:** Who were some of the people you funded?

**Dorothy:** Jerzy Neyman, Harold Hotelling, Betty Scott, Joe Kruskal, Z. W. (Bill) Birnbaum, Allen Wallis, Al Bowker,[15] and all of the other major researchers of the time were funded by ONR. That was before the NSF started. When NSF was started, the man who headed the Mathematics Branch at ONR moved to NSF and headed the Mathematics and Statistics program there. The first thing he did was duplicate the portfolio of our ONR contracts, so all my contractors got double funding. He took away the uniqueness of the ONR funding in basic research in statistics.

**Wendy:** So, the ONR portfolio really gave birth to the NSF program?

**Dorothy:** That is correct. NSF was based on the ONR model.

## YEARS AT THE NATIONAL CENTER FOR EDUCATION STATISTICS AND THE NATIONAL ACADEMY OF SCIENCES

**Dorothy:** While I was at ONR, I was nominated for the Federal Women's Award. At that time, it was given to the five top women in government each year, and I received it. I went to a very elegant dinner in honor of the awardees, and the Assistant Secretary for R&D (research and development) was my escort. He was a very famous man and is still around, but I cannot remember his name. That was a very nice honor. A group of recipients of the award continue to meet twice a year. After ONR, I went to the Office of Education and ran the National Center for Education Statistics. I did a lot of new things there. I started a series of longitudinal studies, which are continuing to this day. I started a fast response survey, so we could get data on policy issues on a timely basis. I did a lot of federal and state relations work because it was extremely important to get the good will of the state people so they would respond to our surveys. It was really hard work. (I remember a man from Texas saying he was sent to meetings to kill as many surveys as possible.) I also went to a lot of meetings of the chief state school officers, because I thought I could be more effective working at that level than working with the people who were responding to the surveys. That was a lot of work because many of the chiefs were political appointees, so every year I had about a dozen new ones to convince of the importance of the surveys. I enjoyed NCES very much. That was a very demanding job, actually. It was hard for my husband, because I traveled so much.

**Ed:** What was your next move after the National Center?

**Dorothy:** I got caught in some nasty politics at the National Center for Education Statistics. In any

---

[15]Jerzy Neyman (1894–1981) was born in Russia, and he was one of the founders of modern statistics. He came to Berkeley in 1939 to create a statistical center within the mathematics department. He remained at the University of California until his death.

Elizabeth Scott (1917–1988) obtained a Ph.D. in astronomy. She had a long collaboration with Jerzy Neyman. Neyman introduced her to statistics, and she introduced him to astronomy.

Joseph B. Kruskal (1929–) was born in New York. He is known for his work in formulating multidimensional scaling and for developing Kruskal's algorithm for computing the minimal spanning tree.

W. Allen Wallis (1912–1998) was an economist and statistician who taught at Yale, Columbia and Stanford Universities. He served as director of research for the Statistical Research Group from 1942 to 1946, after which he joined the faculty of the University of Chicago Graduate School of Business.

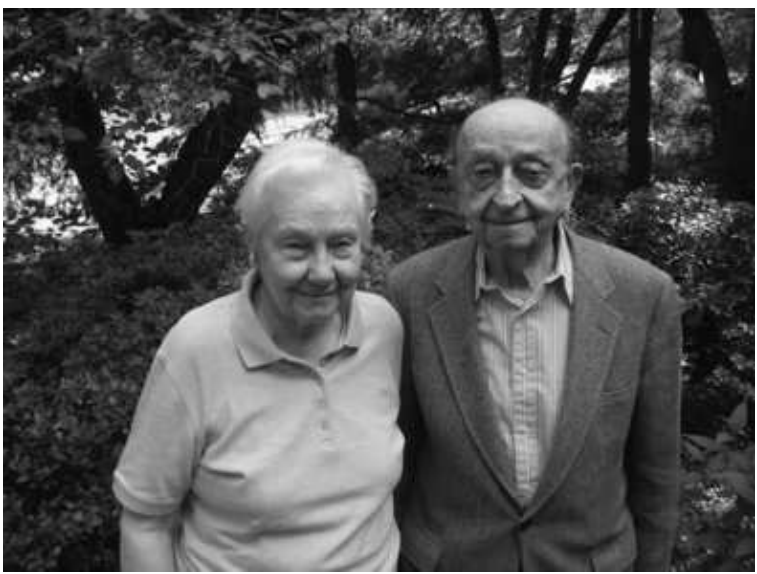

Fig. 7. *This is Dorothy and her husband Leon in May 2006.*

event, I decided I would resign. I was eligible to retire from government service, so I retired.

**Ed:** That would have been roughly when?

**Dorothy:** That was 1974. It was really nasty politics. It involved enmity between people in the Office of the Commissioner and the Department, and I was caught in the middle. I went to work at the National Academy of Sciences. I knew about the job at the Academy in an odd way. I was at one point the Chair of the Conference Board of Mathematical Sciences. At that time, Truman Botts was the secretary for the group, and he had an office in the Academy building. I went to see him occasionally before meetings of the Board.

**Wendy:** Wasn't this called the Board of Mathematical Sciences (BOMS)?

**Dorothy:** No, this was not an Academy entity. This was the organization of all of the mathematics societies. I served on that for the ASA (American Statistical Association) for a while, and then I became an officer. Then I became the Chair. I was the Chair for two years. That's when I interacted with Truman Botts and learned about the job at the Academy. I thought this was a likely place for me to go. I had administered contracts for the National Center for Education Statistics, so many avenues were closed to me. This included Westat, which is a place I would like to have gone, but I had just signed a big contract with Westat. I did, however, get a job at the National Academy as Director of their Human Resources Studies. I worked on that for a while. Later I moved to the Committee on National Statistics and did several studies for the Committee.

**Ed:** So, you were a staff person for CNSTAT?

**Dorothy:** Yes, they are under the Commission of Behavioral and Social Sciences.

**Ed:** The other is the Board of Mathematical Sciences and Applications. When it was first set up it was BOMS.

**Dorothy:** I was always interested in statistics and education. I read a report of a state governors' meeting. They expressed concern about mathematics and science education in this country. They cited studies done by the International Association for Education (IEA). I had seen those studies. I was upset, because they were very flawed studies. The IEA hadn't done anything about nonresponse. They hadn't done anything to ensure the tests that were given to the kids were administered in an equal manner across countries. They hadn't done anything to ascertain whether the questions on the test fairly represented the curricula in the countries, and so forth. It concerned me that the governors were making national policy based on flawed data. I talked to people in the Department of Education and people in the National Science Foundation and came up with the idea of having a Board on International Comparative Studies in Education, which would monitor those studies. The Board would also see to it that future studies were of better quality. The funding was forthcoming, and I ran that Board for around 12 years.

**Ed:** Was that based in the Connecticut Avenue office?

**Dorothy:** No, we were still in Georgetown. The Academy convened a large meeting a few months ago for the dissolution of the Board that I founded. It really wasn't needed anymore, since the Board had done its work. The IEA now has high standards for their studies. Also, the management of the U.S. part was moved to Boston University, where Al Beaton and some other very competent people are involved. The Board didn't need to monitor study quality any longer.

**Wendy:** So, you saw a need, did something about it, and saw it come to fruition?

**Dorothy:** Yes, we monitored the big study called TIMSS from the beginning to the end. That is the Trends in International Mathematics and Science Study. It is widely quoted now. At least one can have confidence in the data.

**Ed:** So you went to the National Academy in 1975. And you were there 19 years.

**Dorothy:** Yes, I decided to retire to spend more time with my husband. He had some health problems and I wanted to have time with him. I'm really glad I did. I did some consulting for a while, and I did some writing.

**Wendy:** I suppose your husband has been very supportive of you—with all of the traveling?

**Dorothy:** He certainly has. Leon is a very patient man, in that regard.

**Wendy:** Well, we've taken up enough of your time. Thank you very much for describing your long and fruitful career in statistics.

**Ed:** Yes, thank you. You can be very proud of your many accomplishments.

## ACKNOWLEDGMENTS

The authors would like to thank June Morita and Peter Guttorp for suggesting this interview and article, George Casella for his encouragement, Ed George for his helpful comments, and Leon and Dorothy for their wonderful hospitality and warmth.